%% file: conference_101719.tex
\documentclass[conference]{IEEEtran}
\IEEEoverridecommandlockouts
\usepackage{cite}
\usepackage{amsmath,amssymb,amsfonts}
\usepackage{algorithmic}
\usepackage{graphicx}
\usepackage{textcomp}
\usepackage{xcolor}
\usepackage{multirow}
\usepackage{multicol}
\usepackage{booktabs}
\usepackage{makecell}
\usepackage{tabularx}
\usepackage[breaklinks]{hyperref}
\usepackage{cleveref}
\usepackage{orcidlink}

\begin{document}
\bstctlcite{IEEEexample:BSTcontrol}

\title{Integrating Energy Efficiency into Software Development: Developer Perspectives and Requirements\\
}

\author{
\begin{tabular}{ccc}
\begin{tabular}[t]{c}
Anika Hennig \orcidlink{0009-0001-9418-8512}\\
\textit{Institute for Digitalization Aachen}\\
\textit{FH Aachen University of Appl. Sci.}\\
Aachen, Germany
\end{tabular}
&
\begin{tabular}[t]{c}
Leonie Schulte \orcidlink{0009-0001-7738-1353}\\
\textit{Institute for Digitalization Aachen}\\
\textit{FH Aachen University of Appl. Sci.}\\
Aachen, Germany
\end{tabular}
&
\begin{tabular}[t]{c}
Philipp M. Zähl \orcidlink{0000-0003-3302-4415}\\
\textit{Institute for Digitalization Aachen}\\
\textit{FH Aachen University of Appl. Sci.}\\
Aachen, Germany
\end{tabular}
\\[5em]
\multicolumn{3}{c}{
\begin{tabular}[t]{c}
Mathias Eggert \orcidlink{0000-0002-3340-7873}\\
\textit{Institute for Digitalization Aachen}\\
\textit{FH Aachen University of Appl. Sci.}\\
Aachen, Germany
\end{tabular}
}
\end{tabular}
}

\maketitle

\begin{abstract}
In the context of the growing energy footprint of information and communication technology, industry optimization efforts have primarily focused on hardware, while the impact of software on energy consumption is often overlooked. Although technical approaches for optimizing software energy consumption have been developed in research, their adoption in everyday development practice remains limited. This study investigates how software developers perceive energy efficiency in their daily work and which requirements and barriers they formulate for AI-assisted tools supporting energy-aware development. As part of the European GreenCode project, ten semi-structured interviews with professional software developers were conducted and analyzed using qualitative content analysis following Mayring’s methodology. The identified requirements were subsequently partly interpreted through the lens of the Technology Acceptance Model. 
The results indicate that energy efficiency rarely plays an explicit role in daily development activities. Instead, energy savings are typically achieved indirectly through performance optimization. Identified barriers to the explicit consideration of energy efficiency include limited awareness and a strong focus on timely delivery. The interviews further revealed requirements for practical tool support, such as actionable optimization suggestions and seamless integration into common development environments. Furthermore, the acceptance of AI-assisted optimization tools strongly depends on transparency regarding the use of data, the actual energy savings compared to the energy consumption of the tool itself, and the disclosure of the training data used.
This study contributes a developer-centered perspective on requirements for energy-aware software development tools and provides insights for designing AI-assisted solutions that align with real-world development practices.
\end{abstract}


\input{chapter/01_introduction}

\input{chapter/02_related_work}
\input{chapter/03_methodology}

\input{chapter/04_results}
\input{chapter/05_discussion}
\input{chapter/06_conclusion}

\section*{Acknowledgment}
This work was conducted within the GreenCode project (ITEA Project 23016) and was supported by German Federal Ministry of Research, Technology and Space (BMFTR) under grant number 01IS24070F. The authors would like to thank all interview participants for sharing their time and valuable insights.

\section{Appendix} \label{appendix} All additional materials and appendixes are available at:\\ \href{https://doi.org/10.17605/OSF.IO/CGY7S}{https://doi.org/10.17605/OSF.IO/CGY7S}

\bibliographystyle{IEEEtran}
\bibliography{references}

\end{document}

%% file: chapter/01_introduction.tex
\section{Introduction}
The energy footprint of information systems (IS) continues to grow, intensifying the need to reduce resource consumption and environmental impact \cite{kile_energy_2025, balanza-martinez_tactics_2024}. While discussions of energy efficiency often emphasize hardware-related aspects, comparatively less attention has been paid to the role of software, even though it contributes significantly to overall energy consumption \cite{ardito_understanding_2015, bozzelli_systematic_nodate, georgiou_software_2020, wysocki_methods_2025}. Unlike hardware, its energy implications are embedded in design and implementation decisions and are therefore less directly observable. This makes energy efficiency particularly challenging to address systematically within everyday software development practice.

In response, research has proposed numerous methods for analyzing and improving the energy behavior of software systems \cite{ardito_understanding_2015, georgiou_software_2020, balanza-martinez_tactics_2024}. However, studies report limited industrial adoption and difficulties in integrating energy-related approaches into everyday development workflows \cite{balanza-martinez_tactics_2024, belgaid_green_2023, guldner_development_2024}. Empirical findings further indicate that awareness, knowledge, and competing development priorities influence whether energy efficiency is considered in practice \cite{ardito_understanding_2015, pereira_ranking_2021, wysocki_methods_2025}. These findings suggest that the existence of technical solutions does not necessarily ensure their sustained use in real-world development contexts. While prior work has examined technical optimization strategies, comparatively less attention has been given to empirically deriving developer-centered requirements for integrating such optimization support into everyday development workflows. 

Recent advances in AI-assisted software engineering provide one possible mechanism for integrating energy-related optimization support into everyday development workflows. Large language models (LLMs) can interpret code context, communicate through natural language, and be embedded into familiar development environments. As a result, they may help make energy-related optimization knowledge more accessible by identifying potential optimization opportunities, explaining their implications, and recommending alternative implementations \cite{vartziotis_carbon_2024, gupta_advancing_2025}. Consequently, the potential of AI lies not in assuming technical superiority, but rather in its capacity to provide contextual support within existing workflows \cite{tuttle_can_2024, rosas_should_2024, gupta_advancing_2025, rasheed_large_2025}.
At the same time, AI-assisted optimization must be considered critically. Current LLM-based development tools do not inherently account for the energy efficiency of generated code, and their operation itself consumes energy and resources \cite{sikand_generative_2024, tuttle_can_2024, vartziotis_carbon_2024, ilager_green-code_2025, rasheed_large_2025}. Consequently, AI-assisted optimization is only environmentally justifiable if its additional resource consumption is outweighed by measurable improvements in the optimized software. 
Against this background, this study investigates an AI-assisted optimization tool as a proposed form of developer-facing support. Rather than evaluating a specific implementation, the study focuses on identifying requirements and barriers that should inform its design and responsible use.

Accordingly, the study seeks to answer the following research questions:

\textbf{RQ1: What is the current role and understanding of energy efficiency in software development?} \\
\textbf{RQ2: What requirements and barriers do developers perceive regarding the integration of an AI-assisted energy efficiency optimization tool into software development workflows?}

Given the focus of RQ2 on the adoption of AI-assisted optimization tools, the Technology Acceptance Model (TAM) \cite{Davis1986TAM} is used as an additional theoretical lens for interpreting the identified requirements and barriers.

The study's results contribute to IS research in three ways: First, it advances theoretical understanding of how energy efficiency is perceived and enacted within software development practice as a socio-technical phenomenon shaped by cognitive, workflow, and organizational factors. Second, it provides empirical insights into developers' perceptions of AI-assisted energy optimization tools, highlighting key barriers such as lack of trust, limited explainability, and concerns about workflow integration. Third, the study provides requirements for integrating AI-based energy optimization tools into software development processes, providing empirically grounded requirements that can inform future design principles.
The paper is structured as follows. Chapter \ref{related_work} provides relevant related work, chapter\ref{methodology} comprises the research design of the study, in chapter \ref{results} all results are presented, in chapter \ref{discussion} we discuss our findings and limitations and in chapter \ref{summary} we provide further research avenues for IS research.

%% file: chapter/02_related_work.tex
\section{Related Work} \label{related_work}

Research on sustainable information technology has traditionally concentrated on the efficiency of hardware systems. More recently, however, both academia and industry have increasingly emphasized the role of software in contributing to the overall energy footprint of IT \cite{ardito_understanding_2015, bozzelli_systematic_nodate, georgiou_software_2020, wysocki_methods_2025}. This development is motivated by evidence that the energy demand of software systems continues to grow at a rapid pace, in some cases exceeding the consumption of entire industrial sectors such as aviation \cite{kile_energy_2025, balanza-martinez_tactics_2024}. Recent surveys and reviews further underline that software-level optimization has emerged as a key lever for improving energy efficiency in ICT systems, particularly in cloud and data center environments \cite{balanza-martinez_tactics_2024, georgiou_software_2020}.

Despite this growing awareness, green software development presents a number of challenges. One challenge relates to the complexity of software systems and the need to balance competing concerns among stakeholders. The adoption of energy-efficient design and development practices often requires a shift in mindset, as developers are accustomed to prioritizing performance, functionality, or maintainability over energy consumption \cite{ardito_understanding_2015}.  They frequently lack awareness of energy efficiency, have little knowledge of suitable practices, and are often uncertain about how software consumes energy \cite{pereira_ranking_2021, wysocki_methods_2025}. Although a variety of tools and good practices are available, they are not yet integrated into comprehensive frameworks that provide developers with a consistent perspective on sustainable software design \cite{ardito_understanding_2015}. Additionally, industrial adoption of green software practices remains low. Prior research indicates that most energy-efficient software techniques are evaluated primarily in academic settings, with limited validation in industrial environments, contributing to slow adoption in practice \cite{balanza-martinez_tactics_2024, guldner_development_2024}.

One reason for this lies in the limited generalizability of research results. In contrast to performance optimization, the energy behavior of algorithms depends to a considerable extent on the characteristics of the execution environment and hardware configuration \cite{belgaid_green_2023}. Empirical studies demonstrate that programming languages, compilers, frameworks, and libraries can lead to different energy consumption profiles \cite{garcia-mireles_interactions_2018, wysocki_methods_2025, pereira_ranking_2021, lange_energy_2025, taha_khudher_green_2024}. For example, Java and Go have been shown to offer similar performance and efficiency, while C\# resulted in higher energy usage in one comparative study \cite{taha_khudher_green_2024}. As a result, attempts to formulate general theories have been limited, and insights from academic studies often risk losing relevance quickly due to the pace of change in software and hardware technologies \cite{belgaid_green_2023}.

Within the research literature, several strategies for improving energy efficiency have been proposed. These approaches are often categorized into code-level refactoring techniques and adaptive runtime strategies, both of which can contribute to energy savings when supported by appropriate profiling and monitoring tools \cite{ardito_understanding_2015, moises_practices_2018}. Refactoring methods seek to reduce energy usage by removing patterns or instructions that lead to higher consumption, often applied during maintenance where they also contribute to code quality \cite{ardito_understanding_2015, moises_practices_2018, manimegalai_energy_2024}.  
Self-adaptive approaches, in contrast, aim to enable software systems to select between different configurations depending on contextual factors, thereby adjusting their energy profile dynamically \cite{ardito_understanding_2015}. Both approaches can be applied iteratively, with improvements verified using power profiling tools.

At the implementation level, a number of green coding practices have been documented. These include the use of efficient algorithms and data structures, minimizing resource usage, reducing unnecessary computations, and optimizing memory management \cite{gupta_advancing_2025, kile_energy_2025, salmikuukka_guidelines_2024}. Practical recommendations include buffering I/O operations, avoiding polling, and using language features that require fewer computational resources, for example, the use of for-loops instead of while-loops in Python, or the adoption of standard libraries in place of handwritten code \cite{procaccianti_empirical_2016, cappendijk_generating_2024}. Beyond individual practices, several authors emphasize that integrating energy considerations across all phases of the software development lifecycle (SDLC) leads to more sustainable outcomes \cite{balanza-martinez_tactics_2024, georgiou_software_2020, gupta_advancing_2025, kile_energy_2025, kruglov_developing_2023, tiwari_evaluating_2023}. Agile development methods have also been identified as suitable for supporting green software practices \cite{wysocki_methods_2025}.

The measurement and evaluation of software energy efficiency remains a key research focus. While there is broad agreement on the importance of appropriate criteria and metrics, a standardized set of measures has not yet emerged. Metrics currently employed include CPU, GPU, memory and network usage, I/O operations, runtime, floating-point operations, and carbon emissions \cite{kile_energy_2025, bozzelli_systematic_nodate, ergasheva_metrics_2020, garcia-mireles_interactions_2018, moises_practices_2018, ilager_green-code_2025, kruglov_developing_2023, rajput_enhancing_2024, vartziotis_learn_2024, vartziotis_carbon_2024}. 

To obtain these measurements, both hardware- and software-based methods are employed. Hardware-based monitoring, such as external meters or specialized platforms like GreenMiner, EET, or SEFLab, provides accurate system-level data but is often difficult to scale \cite{lange_energy_2025, georgiou_software_2020, wysocki_methods_2025, procaccianti_empirical_2016, rajput_enhancing_2024}. Software-based approaches rely on interfaces such as RAPL and include tools like PowerTOP, Perf, Joulemeter, and glcb \cite{rajput_enhancing_2024, lange_energy_2025, ardito_understanding_2015}. Integration into development workflows has also been explored, with frameworks such as ENRICO providing feedback within CI/CD pipelines, and tools like J-Referral offering guidance on energy-efficient JVM configurations \cite{lange_energy_2025, belgaid_green_2023}. Despite the availability of numerous measurement tools, several studies report that reproducibility, accuracy, and representativeness remain persistent challenges with most solutions being difficult to integrate into everyday development workflows, limiting their practical usefulness for developers \cite{belgaid_green_2023, lange_energy_2025}. While containerization has been proposed as one way of addressing reproducibility, hardware variability and background processes continue to complicate measurement accuracy \cite{belgaid_green_2023, rajput_enhancing_2024, guldner_development_2024, balanza-martinez_tactics_2024, procaccianti_empirical_2016}.

A more recent development in the field concerns the role of artificial intelligence and LLMs in software development. Studies indicate that LLM-generated code often prioritizes correctness and runtime performance, while energy efficiency is often not explicitly considered \cite{sikand_generative_2024, tuttle_can_2024, vartziotis_carbon_2024}. Research in this area is beginning to examine how LLMs can be guided toward greener code, for example through prompt engineering that incorporates explicit instructions for energy optimization \cite{cappendijk_generating_2024}. Other approaches include the development of frameworks such as GREEN-CODE, which uses reinforcement learning to optimize trade-offs between accuracy, latency, and energy consumption during code generation \cite{ilager_green-code_2025}. Broader integration of AI into the SDLC has also been proposed, with LLMs envisioned as tools to recommend energy-efficient libraries and support continuous optimization \cite{gupta_advancing_2025, paul_comprehensive_2023}. Tools such as GitHub Copilot and Amazon Q already support developers with tasks such as code generation, completion, and refactoring \cite{ilager_green-code_2025, rasheed_large_2025, tuttle_can_2024, vartziotis_learn_2024}. Nevertheless, the computational overhead of AI-based approaches raises concerns about their overall sustainability benefits \cite{ilager_green-code_2025, rasheed_large_2025}. 

Taken together, the state of research highlights progress in conceptual frameworks, strategies, and measurement techniques, but also reveals important limitations. The field remains fragmented, industrial adoption is limited, and findings are often difficult to generalize. Future research is therefore directed toward the development of standardized metrics and tools, the integration of energy-awareness across all phases of the SDLC, and closer collaboration between academia and industry. Empirical validation in industrial settings is regarded as essential for ensuring that proposed solutions are representative and applicable in practice \cite{balanza-martinez_tactics_2024, guldner_development_2024, garcia-mireles_interactions_2018, tiwari_evaluating_2023, kile_energy_2025}. In this context, greater attention to practitioner perspectives is necessary, particularly regarding the requirements, barriers, and acceptance factors associated with integrating AI-assisted energy optimization into existing workflows. This study addresses these gaps by empirically examining developers’ current understanding of energy efficiency and their perspectives on AI-assisted optimization support.

%% file: chapter/03_methodology.tex
\section{Methodology} \label{methodology}

To address the research questions, a qualitative research design was adopted. The study examines how software developers describe the role and understanding of energy efficiency in their everyday work and how they perceive requirements and barriers related to the integration of AI-assisted optimization tools. Both research questions concern interpretations, experiences, and evaluations situated within professional practice.

A qualitative approach was considered appropriate because the study seeks to understand how developers interpret and articulate their experiences within their everyday work contexts. As \cite{BrinkmannKvale2018DoingInterviews} states, qualitative interviewing makes it possible to “understand, describe, and sometimes explain social phenomena ‘from the inside’.” This perspective is particularly relevant for examining how energy efficiency is currently positioned within development practice, which requirements developers formulate regarding optimization tools, and how AI-based support is assessed, including perceived benefits and concerns.

Data was collected through semi-structured interviews. This format provides a clear thematic structure and encourages participants to speak openly about their experiences and perspectives. It ensures that key topics derived from the research questions are addressed across interviews, yet leaves sufficient flexibility to pursue relevant aspects that emerge during the conversation. \cite{Adams2015ConductingSemiStructuredInterviews} The interview guide (see Table \ref{tab:int_gui}) consisted of open-ended questions derived from and structured along the research questions, following the recommendations of \cite{BrinkmannKvale2018DoingInterviews}. In this process, both research questions were broken down into several smaller questions formulated in simpler language, enabling the collection of relevant information to address the research objectives. At the same time, the semi-structured format permitted additional follow-up questions when clarification or further exploration was needed.

\begin{table}[htbp]
\caption{Interview Guide (translated from German to English)}
\label{tab:int_gui}
\begin{tabularx}{\linewidth}{|X|p{12em}|}
\hline
\multicolumn{2}{|p{\dimexpr\linewidth-2\tabcolsep-2\arrayrulewidth\relax}|}{\textbf{Opening} Greeting the interviewee and thanking them for their time. Brief introduction of the topic and explanation of the procedure. Reference to confidentiality and anonymity of the data. Request for openness during the interview. Clarification of consent to audio recording. Note that anonymized excerpts may be used in scientific publications. Introduction of the GreenCode project: ``In the GreenCode project, we develop AI-supported methods to identify inefficient program code in software-based systems and improve its energy efficiency, both in existing systems and in new developments. The goal is to reduce energy consumption in IT landscapes without compromising functionality or security. The project is carried out by an international consortium of 26 institutions and companies from nine countries. The present interviews are conducted as part of the requirements analysis to derive practical requirements for such tools.''} \\
\hline
\textbf{Question} & \textbf{Background} \\
\hline
Age, current position \& responsibilities, company size 
& Enables interview comparisons filtered by demographic characteristics and provides transparency of the sample \\
\hline
Which development tools, frameworks, or technologies do you regularly use in your daily work? & Captures the current state of practice and helps identify potential integration points for future solutions. \\
\hline
Are there specific tools or utilities you use to analyze or optimize your code? & Provides insight into existing routines and practices related to performance or efficiency improvements. \\
\hline
How present is the topic of energy efficiency in your software development processes? & Explores the relevance and visibility of energy efficiency within everyday development activities. \\
\hline
What does sustainable software mean to you personally - what aspects does it include? & Examines how developers conceptualize sustainable software \\
\hline
What obstacles do you see when aiming to develop software in a more sustainable or energy-efficient way? & Identifies technical, organizational, or process-related barriers to sustainable software development. \\
\hline
Imagine you had a tool that helps you write more energy-efficient code - what kind of feedback or information would be helpful to you? & Explores expectations regarding feedback types, level of detail, and actionable guidance. \\
\hline
What additional capabilities would such a tool need, or how should it function, for you to use it regularly in your daily work? & Investigates requirements related to integration, workflow compatibility, and ease of use. \\
\hline
In our project, we are specifically considering the use of AI - what potential do you see in this context? & Examines perceived benefits and expectations associated with AI-supported development tools. \\
\hline
Is there anything that would discourage you from using AI-based tools? & Identifies potential concerns or limiting factors affecting acceptance and adoption. \\
\hline
From your perspective, is there anything important that we have not yet addressed? & Provides space for additional perspectives, insights, or topics not covered in the structured questions. \\
\hline
\end{tabularx}
\end{table}

\subsection{Sample}
A total of 10 interviews were conducted in this study (see Table \ref{tab:participants}). This number lies within the typical range of approximately 15 (±10) interviews reported for qualitative interview research \cite{BrinkmannKvale2018DoingInterviews}. Sample sufficiency was assessed retrospectively based on the recurrence of the main categories across the final interviews.

All participants were software developers at the time of the interviews, with one participant (I04) working as a research associate but with a background in software development and continued involvement in software-related activities. The participants had between 2 and 23 years of professional experience working in software teams, with an average of 7.5 years. The sample consisted of nine male participants and one female participant. Given the limited gender diversity of the sample, the study does not permit meaningful conclusions regarding potential gender-related differences in perceptions or requirements.

The interviewees were employed in organizations of different sizes: four participants (I02, I06, I07, I09) in large companies, four participants (I03, I05, I08, I10) in medium-sized companies, one participant (I01) in a micro enterprise, and one participant (I04) at a university.

The interviews lasted between 11 and 35 minutes, with an average duration of 22 minutes. Each interview addressed 10 guiding questions in addition to demographic questions. This corresponds to the guideline of approximately 8–15 questions per hour \cite{BrinkmannKvale2018DoingInterviews}. 

Participants were recruited through the authors’ respective professional networks. Prior to the interviews, participants were informed about the purpose and scope of the study as well as the voluntary nature of their participation and the handling of the collected data. All interviews were conducted with informed consent. To ensure confidentiality, transcripts were anonymized and any identifying information, such as names, companies, or project references, was removed or generalized. The anonymized and translated interview transcripts, as well as the detailed tool landscape of the participants, are available via the Open Science Framework (OSF), referenced in the \hyperref[appendix]{appendix}).

\begin{table}[htbp]
\caption{Overview of Interview Participants}
\label{tab:participants}
\begin{tabularx}{\linewidth}{|X|p{2em}|p{3em}|X|X|X|X|X|}
\hline
\textbf{Interviewee} & \textbf{Age} & \textbf{Gender} & \textbf{Job} & \textbf{Company Size} & \textbf{Experience (Years)} \\
\hline
I01 & 28 & m & Software Developer & Micro  & 14 \\
\hline
I02 & 29 & m & Software Developer & Large  & 5  \\
\hline
I03 & 26 & m & Software Developer & Medium & 3  \\
\hline
I04 & 29 & m & Research Associate (former Software Developer) & Large  & 2.5 \\
\hline
I05 & 31 & m & Software Developer & Medium & 12 \\
\hline
I06 & 28 & m & Software Developer & Large  & 3  \\
\hline
I07 & 43 & m & Software Developer & Large  & 23 \\
\hline
I08 & 23 & f & Software Developer & Medium & 4  \\
\hline
I09 & 30 & m & Software Developer & Large  & 6  \\
\hline
I10 & 25 & m & Software Developer & Medium & 2  \\
\hline
\end{tabularx}
\end{table}

\subsection{Data Collection}
Data collection was conducted through ten semi-structured interviews. The interviews were carried out by a single person. Nine interviews took place remotely via video chat, while one interview (I01) was conducted in person.

Prior to the main data collection, one in-person pilot interview was conducted. The pilot served to review the interview guide with regard to clarity, structure, and anticipated duration \cite{Adams2015ConductingSemiStructuredInterviews}.

All interviews were audio-recorded with prior informed consent. A simple transcription approach was applied, as the analysis focused on the verbal content of the interviews rather than on detailed speech characteristics \cite{DresingPehl2020Transkription}.

The recordings were transcribed using \href{https://www.microsoft.com/en-us/microsoft-365/}{\textit{Microsoft 365 Transcribe feature}}. As automated transcription may produce inaccuracies, the transcripts were subsequently processed with the assistance of \href{https://openai.com/chatgpt/overview/}{\textit{ChatGPT-4o mini}} to correct evident transcription errors and formatting inconsistencies. The model was explicitly instructed not to paraphrase, reformulate, or modify sentence structures. Its use was restricted to identifying inaccuracies or incomplete fragments generated during automated transcription.

Following this step, all AI-assisted revisions were manually compared with the original audio recordings to prevent potential errors or distortions and to ensure that wording and meaning remained unchanged \cite{Taylor2024ChatGPTTranscriptions, Hayes2025ConversingLLMs}. During the review process, identifying information was removed or generalized to ensure anonymity. Only the fully anonymized and manually verified transcripts were used for the subsequent analysis.

\subsection{Data Analysis}
The transcribed interviews formed the basis of the analysis. The data is analyzed using structuring qualitative content analysis following Mayring’s methodological framework \cite{Mayring2022QualitativeInhaltsanalyse}. The objective was to systematically examine how developers describe the role and understanding of energy efficiency in software development and how they articulate requirements and barriers regarding AI-assisted optimization tools.

Each interview was first read in full to gain an overall understanding of the material before relevant text passages were identified and coded using \href{https://www.maxqda.com/}{\textit{MAXQDA}}. The coding process began openly, allowing codes to be developed directly from the interview data.

In subsequent steps, related codes were grouped and condensed into categories. Through repeated comparison across interviews, categories were differentiated, merged, or reorganized where necessary. This process resulted in a structured category system consisting of seven main categories with corresponding subcategories.

A coding guideline was established to document the category system. For each category, a category name, definition, coding rules, and anchor examples were specified to ensure consistent application. The guideline was iteratively refined as the category structure stabilized. The full coding guideline, including the category system, can be found in the \hyperref[appendix]{appendix}.

In a later stage of the analysis, the emerging category system was interpreted with reference to the TAM \cite{Davis1986TAM}. As the interview guide included questions focused on expected benefits, feedback mechanisms, and conditions for regular adoption, the TAM constructs \textit{Perceived Usefulness} and \textit{Perceived Ease of Use} provided a suitable conceptual lens for structuring and interpreting the identified requirements. Coding decisions and category assignments were reviewed to ensure internal consistency.

To assess the reliability of the coding process, an intercoder reliability analysis was conducted. Two researchers independently coded the interview material using the developed coding guideline to assess intercoder agreement. Agreement was calculated as proportion agreement based on overlapping coded segments in MAXQDA.
Initial differences in coding were reviewed and discussed within the research team in order to clarify category definitions and refine the coding guideline. This iterative discussion of disagreements is recommended in qualitative research as a means of improving the consistency and precision of category systems \cite{RaedikerKuckartz2019Intercoder, MacPhail2016QualitativeIntercoderReliability, Campbell2013CodingReliability}. After revising the coding rules and resolving remaining disagreements through negotiated agreement \cite{Campbell2013CodingReliability}, the intercoder agreement reached 74.85\%, which lies within the range commonly considered acceptable for exploratory qualitative studies \cite{Campbell2013CodingReliability, MacPhail2016QualitativeIntercoderReliability}.

%% file: chapter/04_results.tex
\section{Results} \label{results}

The results are organized along the central topics that emerged from the interviews, progressing from the current state of practice to design implications for AI-assisted optimization tools. First, we examine the existing role of energy efficiency within companies and how participants personally conceptualize sustainable software. Building on this understanding, we then address barriers that hinder its adoption in practice. Subsequently, we turn to requirements for tool support, distinguishing between functional expectations and interaction-related aspects. Finally, we explore the perceived benefits and barriers associated with integrating artificial intelligence into such tools. Together, these sections reflect both organizational and technical perspectives identified in the data. The corresponding categories and their definitions, summarized in \cref{tab:subcategories_pres_ener_effi,tab:subcategories_pers_und_sus_sof,tab:subcategories_barr_sus_sof,tab:subcategories_perc_use,tab:subcategories_perc_ease_use,tab:subcategories_bene_ai,tab:subcategories_barr_ai}, were developed inductively from the interview material.

\subsection{Presence of energy efficiency in the companies}
For most interviewees (I01, I02, I04-I10), energy efficiency does not play an explicit role in everyday software development practice. Only one participant (I03) reported that energy efficiency is actively addressed within development activities due to the presence of a dedicated focus group within the company investigating the energy consumption of different programming languages.
Across the remaining interviews (I01, I02, I04-I10), energy efficiency was either not considered in software development (I01, I07) or only addressed implicitly (I02, I04, I08, I10). In several cases, participants emphasized that development activities primarily focus on functionality and overall performance, with energy efficiency emerging merely as a byproduct of these priorities. For example, it was associated with performance optimization at the hardware level (I04, I10) or with general practices such as resource-saving approaches and clean code principles, rather than being pursued as an explicit energy-related objective (I02, I08).
Some participants additionally noted that sustainability is recognized as a relevant topic at the organizational level, particularly within management (I06, I09) or due to the company’s industry context (I05). However, its fulfillment in software development remains limited. 

\begin{table}[htbp]
\caption{Presence of energy efficiency in the companies}
\label{tab:subcategories_pres_ener_effi}
\begin{tabularx}{\linewidth}{|p{6em}|X|p{4.5em}|}
\hline
\textbf{Subcategory} & \textbf{Definition} & \textbf{Interviewee} \\
\hline
Present & Energy efficiency is known and is specifically addressed in the company & I03 \\
\hline
Not present & Energy efficiency is not a topic and is not considered in the company &  I01, I07 \\
\hline
Energy savings through performance focus & By focusing on performance and targeted resource use, energy is saved, even though this is not the primary goal & I02, I04, I08, I10 \\
\hline
Present in the company, but not in SD & Energy efficiency is a relevant topic in the company, but it is not considered in software development & I05, I06, I09 \\
\hline
\end{tabularx}
\end{table}

\subsection{Personal understanding of sustainable software} 
All interviewees associated sustainable software primarily with resource efficiency. Resource efficiency was commonly described as reducing resource consumption through efficient programming and optimized system utilization (I02, I04-I06, I09, I10). Examples mentioned by participants included loading only components that are actually required (I02, I08, I09) and using servers and computing resources only when needed (I01, I09). In this context, sustainable software was further linked to improving execution efficiency in order to limit energy consumption at runtime (I03, I07).
Beyond resource efficiency, participants associated sustainability with long software lifespans, which participants linked to clear code structures, comprehensive documentation, and traceability (I03, I04, I08). Additionally, an interviewee highlighted the deliberate selection of tools as a sustainability factor, noting that certain frameworks or programming environments may introduce unnecessary background processes and thereby increase energy consumption. The programming language Java in combination with the Spring Framework was mentioned as an example of such behavior (I02).
Finally, a participant extended the definition of sustainable software to include social and societal aspects. In this perspective, software sustainability was described as also involving the minimization of potentially harmful usage patterns, such as the manipulation of vulnerable users through content algorithms on social media platforms (I06).

\begin{table}[htbp]
\caption{Personal understanding of sustainable software}
\label{tab:subcategories_pers_und_sus_sof}
\begin{tabularx}{\linewidth}{|p{6em}|X|p{4.5em}|}
\hline
\textbf{Subcategory} & \textbf{Definition} & \textbf{Interviewee} \\
\hline
Resource Efficiency & Software is designed and implemented in such a way that only the resources that are actually necessary are used during its execution, in order to avoid unnecessary energy and resource consumption & I01-I10 \\
\hline
Long-lasting software & The software is designed for a long service life, e.g., through good maintainability, comprehensive documentation, high traceability, and clear structure, to save resources for new developments & I03, I04, I08 \\
\hline
Conscious use of tools & The impact of tools, frameworks, and architectures on energy consumption is considered when making selections & I02 \\
\hline
Social aspects & The software is designed and used in a way that has no harmful effects on individuals or society & I06 \\
\hline
\end{tabularx}
\end{table}

\subsection{Barriers to implementing sustainable software}
First, participants (I03-I06) described limited awareness and knowledge regarding sustainable software development as an obstacle to the adoption of sustainable practices. It was emphasized that existing knowledge gaps must first be addressed before sustainable practices can be effectively implemented in companies (I05, I06, I10). These knowledge gaps were closely linked to a lack of awareness, as developers often perceive software development as having only a minor impact on overall carbon emissions (I03, I04). Consequently, sustainability was considered less relevant unless developers personally perceive its impact or gain a deeper understanding of its broader implications (I03).

Next to knowledge and awareness gaps, most interviewees identified the preeminence of functional requirements and time pressure over sustainability objectives as a central barrier to sustainable software development (I01, I02, I04, I05, I08-I10). Participants frequently referred to the pressure created by upcoming release deadlines, which leads to a strong focus on delivering functional features and completing software on schedule (I02, I04, I05, I08). Within this context, allocating additional time for sustainability-related optimizations was described as difficult, as such efforts were perceived as time-consuming (I01, I02, I04, I05, I08, I09). A participant further highlighted a reluctance to engage with new sustainability practices, describing a general aversion to additional learning and effort when the software's functionality already meets project requirements (I10).

In addition to time and priority constraints, sustainable practices were also described as technically challenging. As different languages exhibit varying energy consumption characteristics, some participants argued that programming languages themselves introduce inefficiencies (I03, I06). Furthermore, concerns were raised that optimized code, while reducing resource consumption, may increase execution time (I09).

Beyond technical considerations, organizational factors were described as significant barriers. Participants (I04, I05, I07) emphasized that active management support is required to successfully integrate sustainable practices. It was mentioned that without such support, sustainability-related activities may be perceived as an inefficient use of development time (I05). Furthermore, gaining management approval was seen as difficult, particularly because of the participants association that sustainability improvements are not directly billable to the customer and therefore require companies to initially bear the associated costs (I04, I07).

Closely related to these concerns, the expected benefits of sustainable software development were frequently perceived as low compared to the required investment (I01, I07, I09). Participants argued that the visible benefits for customers are limited (I01, I07, I08), which in turn reduces customers’ willingness to pay for sustainability-related improvements (I01, I07). As a result, organizations would often have to bear the additional costs themselves (I01, I07). One participant illustrated this perception by explaining that customers are unlikely to pay more for a product solely because it consumes less energy: 
\begin{quote}
    “It's not present because the effect you get from it, or the benefit, isn't really there. Because it's a bit harsh to say, when I build a web application, it doesn't run on my infrastructure, but on the customer's cell phone or laptop. Well, it's not that important to me that it's energy-efficient, because the customer wouldn't pay for it either.  They wouldn't say, 'I'll pay 20\% more because my cell phone uses less power.' It might make sense, but it still doesn't happen. That means it's difficult to sell.” (I01).
\end{quote}
In addition, personnel costs were described as a barrier, as companies are perceived to prioritize lower-cost developers or solutions that focus on functional completion rather than optimization (I01, I07). Participants emphasized that the development effort required for energy-related optimizations is often considered disproportionate to the potential cost savings. For example, potential reductions in cloud resource usage or storage consumption were described as economically insignificant compared to the developer time required to implement such optimizations (I01).

\begin{table}[htbp]
\caption{Barriers to implementing sustainable software}
\label{tab:subcategories_barr_sus_sof}
\begin{tabularx}{\linewidth}{|p{6em}|X|p{4.5em}|}
\hline
\textbf{Subcategory} & \textbf{Definition} & \textbf{Interviewee} \\
\hline
Lack of awareness and knowledge gaps & Noticeable knowledge and awareness gap exist regarding sustainable software. & I03-I06, I10 \\
\hline
Focus on functionality and timely delivery & Functional requirements and time pressure dominate over sustainability focus. & I01, I02, I04, I05, I08-10 \\
\hline
Technical barriers & Sustainable software development is perceived as technically limited when optimizations cause conflicts of interest between resource consumption and execution time or are restricted by technology-related inefficencies. &  I03, I06, I09 \\
\hline
Management & Sustainable software development must be actively supported by management in order to be successfully integrated into everyday work. & I04, I05, I07 \\
\hline
Insufficient cost-benefit perception & The expected benefits of sustainable software are considered too low for companies and customers, especially in view of the associated costs. & I01, I07-I09 \\
\hline
\end{tabularx}
\end{table}

\subsection{Perceived Usefulness of optimization tool}
While the previous findings describe how developers currently perceive energy efficiency in software development, the perceived usefulness aims to capture developers’ expectations and requirements regarding the use of an energy efficiency optimization tool in their habitual development workflow. 
In this regard, most interviewees expressed a preference for concrete optimization suggestions that are directly applicable in practice (I01-I03, I05-I10). These suggestions were expected to be grounded in prior code analysis to ensure contextual relevance and technical feasibility (I06, I07, I09, I10). In addition, a participant highlighted that the tool should iteratively refine its recommendations based on performance metrics and defined thresholds, thereby progressively improving optimization outcomes (I10). Furthermore, a comprehensive review process after the completion of the code was deemed beneficial (I02).

Beyond the provision of optimization suggestions, participants highlighted the importance of explainability. Explanations accompanying proposed changes were perceived as essential for improving developers’ understanding of sustainability-related issues and for supporting the long-term adoption of sustainable development practices (I01, I03, I04, I07-I10):
\begin{quote}
    “It's not just about making suggestions or proposing improvements, but also about justifying them. That way, as a user, you become more aware of the issue and might automatically incorporate it into your own workflow - so that later on, feedback might not even be necessary anymore because you're already doing it yourself.” (I04)
\end{quote}In that context, it was considered important to present the thresholds used for the optimizations as transparent as possible, to enhance explainability and understanding (I09).

Perceived usefulness was further linked to flexibility and customization. One participant suggested enabling individual configuration through versioning mechanisms or configuration files such as YAML (I01). Additionally, interviewees proposed allowing users to specify optimization goals or requirements directly, for example, through conversational interfaces such as chat-based input (I01, I10). Customizable monitoring interfaces, for example, in conjunction with frameworks like OpenTelemetry, were suggested to allow developers to adapt visualizations to their specific requirements (I01). The provision of APIs was mentioned as a way to integrate the tool into monitoring systems and to enable transparency regarding application performance and tool usage (I01). 

Finally, participants emphasized the importance of making the impact of optimizations transparent (I02-I07, I09). Reports and metrics quantifying effects such as energy consumption, runtime performance, and storage efficiency were considered particularly valuable (I03, I05-I07, I09). Suggested features included direct comparisons between existing and optimized code versions (I04, I09), real-time visualizations such as live monitoring windows displaying energy consumption during execution (I03), and automated reporting to support development and performance testing processes (I06). Such measurable outcomes were also regarded as useful for communicating benefits to management by providing concrete evidence of sustainability improvements (I07).

\begin{table}[htbp]
\caption{Perceived Usefulness of optimization tool}
\label{tab:subcategories_perc_use}
\begin{tabularx}{\linewidth}{|p{6em}|X|p{4.5em}|}
\hline
\textbf{Subcategory} & \textbf{Definition} & \textbf{Interviewee} \\
\hline
Suggestions for improvement & Identifying optimization potential in the code and deriving concrete improvement proposals & I01-I03, I05-I10 \\
\hline
Iterative training & Tool arrives at better solutions iteratively through comparison options and metrics & I10 \\
\hline
Reasoning & Providing justification for the proposed changes by presenting the underlying reasons, the intended specific improvements, and the relevance and necessity of these improvements &  I01, I03, I04, I07-I10 \\
\hline
Configuration & Users should be able to control the tool's settings flexibly and in a versionable manner & I01 \\
\hline
Text field for optimization requests & Free text field for entering individual requests, instructions, or goals for optimization & I01, I10 \\
\hline
Monitoring & User-friendly interface for clear presentation of information such as usage, status, or processes & I01 \\
\hline
Impact & Presenting the effects of proposed changes, e.g., in the form of a report & I02-I07, I09 \\
\hline
\end{tabularx}
\end{table}

\subsection{Perceived Ease of Use of optimization tool}
The perceived ease of use focuses on developers’ design expectations and requirements regarding the optimization tool.
As a central requirement, the design and presentation of feedback emerged (I02-I05, I07, I08, I10). Interviewees expressed a preference for live feedback, meaning that the tool should operate during execution and provide immediate optimization suggestions (I03-I05, I07, I10). Real-time feedback was considered particularly beneficial for maintaining workflow continuity, as it enables developers to address optimization issues directly instead of postponing them (I03, I07). It was emphasized that optimization suggestions should be clearly highlighted within the development environment, for example, through linter-like visual indicators (I02). At the same time, feedback should remain non-disruptive; intrusive elements such as pop-ups were explicitly rejected in favor of more subtle visual cues that do not interrupt the development process (I03, I08). Suggestions for presentation included integration directly within the code editor (I04), display in an additional window (I04), implementation as a browser-based tool (I10), or in the CI pipeline (I10).

Another important aspect of perceived ease of use concerned seamless integration into existing development environments (I02, I04-I08). Participants noted that the better the tool aligns with established workflows, the more likely it is to be adopted in practice (I02, I07). In this context, incorporating AI-based assistance directly within the IDE environment was suggested as a promising approach (I04).

Finally, ease of use was explicitly associated with intuitive design and minimal interaction effort (I04, I09). The tool was expected to require little additional effort to operate and to support regular use without increasing cognitive load (I09). Additional aspects mentioned by individual participants included gamification elements to maintain developer engagement (I08) and the ability to use the tool offline, which was considered important when working without stable internet access (I04).

\begin{table}[htbp]
\caption{Perceived Ease of Use of optimization tool}
\label{tab:subcategories_perc_ease_use}
\begin{tabularx}{\linewidth}{|p{6em}|X|p{4.5em}|}
\hline
\textbf{Subcategory} & \textbf{Definition} & \textbf{Interviewee} \\
\hline
Feedback design & Optimization feedback is provided visually, context-dependend, and in a way that supports the workflow, both during programming or after development. & I02-I05, I07, I08, I10 \\
\hline
Integration into development environments & Integration into existing development environments or other existing tools & I02, I04, I05, I06, I07, I08 \\
\hline
Ease of use & Intuitive and clearly structured operation without significant effort or complex setup & I04, I09 \\
\hline
Gamification & Use of playful elements such as rewards or progress indicators to increase motivation and engagement & I08 \\
\hline
Offline use & Basic functions of the tool are usable without an internet connection & I04 \\
\hline
\end{tabularx}
\end{table}

\subsection{Benefits of AI-assisted optimization tool}
The interview data indicate several perceived benefits of integrating artificial intelligence into the proposed tool. Most prominently, AI was associated with increased development efficiency and speed. Interviewees emphasized that AI can reduce development time by taking over tasks that are often perceived as tedious, such as performance testing (I03, I05, I07-I09). In addition, the direct integration of AI into the tool was regarded as beneficial, as it eliminates the need to switch between external resources, including search engines or standalone AI systems, thereby supporting faster and more efficient problem-solving (I03, I05, I07-I09). AI was further described as versatile, given its perceived applicability to different development tasks and use cases (I10). One participant (I09) highlighted the potential of the quantization of AI models to enable deployment on low-end hardware: 
\begin{quote}
    "Today, quantization can already be used to compress AI models so that they can run on less powerful hardware - in other words, on low-end hardware, in quotation marks. Of course, the result is not as good as with the large professional models from well-known providers, but there is still huge potential there." (I09).
\end{quote}
 
Beyond efficiency gains, AI was described as expanding the scope of optimization through its ability to process large amounts of data (I01, I02, I05, I06). Participants noted that AI can help concretize vague information, identify recurring patterns such as inefficient code structures, and analyze large codebases more effectively than traditional rule-based approaches (I01, I02, I06). When trained on high-quality code examples, AI was further perceived as capable of providing best-practice recommendations (I05). In this regard, personalized feedback and natural-language explanations were valued for supporting both optimization decisions and developers’ understanding (I04, I07).

\begin{table}[htbp]
\caption{Benefits of AI-assisted optimization tool}
\label{tab:subcategories_bene_ai}
\begin{tabularx}{\linewidth}{|p{6em}|X|p{4.5em}|}
\hline
\textbf{Subcategory} & \textbf{Definition} & \textbf{Interviewee} \\
\hline
Efficiency and speed & AI can accelerate work processes and make them more efficient, and can be used in a variety of contexts & I03, I05, I07-I10 \\
\hline
Expanded optimization scope through large data sets & AI uses access to extensive data to identify patterns and relationships that are difficult to detect with traditional methods, making complex or recurring optimization opportunities visible  &  I01, I02, I05, I06 \\
\hline
Personalized feedback & AI adapts feedback to the knowledge level, style, and needs of individual users & I04, I07 \\
\hline
\end{tabularx}
\end{table}

\subsection{Barriers to acceptance of an AI-assisted optimization tool}
The interviewees identified several barriers that constrain the acceptance of AI-based support for energy-efficient software development. A frequently mentioned concern was the insufficient quality and reliability of AI-generated results (I01, I04, I06, I07, I09). Participants described AI systems as prone to errors and hallucinations (I04, I09). Concerns were also raised regarding the quality of training data. As training data is often collected from the internet, where AI-generated content is already present, participants perceived a risk of feedback loops in which incorrect or low-quality outputs are reused for further model training, thereby potentially degrading result quality over time (I07, I09). In addition, limitations in context size were mentioned as problematic, particularly when analyzing large and complex systems that require extensive contextual information (I06).
Closely related to these concerns were issues of trust in AI systems. On the one hand, a lack of trust was attributed to the perceived error-proneness of AI-generated results (I01, I04). On the other hand, some participants warned against excessive trust, expressing concerns that developers might accept results without sufficient verification, which could reduce critical thinking, reflective ability, and deeper understanding of the underlying content (I07, I09).
Another important barrier concerned data protection and data sovereignty, particularly in the European context. Participants noted that meaningful AI-based analysis would likely require access to large portions of source code, which raised concerns about potential leakage of valuable company data or sensitive customer information (I03, I05, I07, I09). These concerns were closely linked to regulatory requirements, such as compliance with the General Data Protection Regulation (GDPR), which could further complicate the practical adoption of AI-based tools (I03, I05, I07, I09).
In addition, several participants expressed concerns regarding the energy and resource consumption of AI systems (I02, I03, I05, I06, I09, I10). The training of AI models was described as particularly resource-intensive, leading participants to suggest that the use of pre-trained models may be more feasible than custom model training (I02). Participants also emphasized that AI systems themselves require considerable computational resources and electricity, in some cases significantly more than simpler tools such as search engines (I02, I09, I10). As a result, the overall benefit of AI-supported optimization was questioned, as the energy savings achieved might be small compared to the energy consumed by the AI itself (I02, I03, I05, I09). Additional concerns included the potential need for more powerful hardware or extended development setups, such as remote servers, which could further increase resource demands (I06).
Finally, broader fears regarding the increasing influence or perceived superiority of AI systems in software development contexts were raised (I09).

\begin{table}[htbp]
\caption{Barriers to acceptance of an AI-assisted optimization tool}
\label{tab:subcategories_barr_ai}
\begin{tabularx}{\linewidth}{|p{6em}|X|p{4.5em}|}
\hline
\textbf{Subcategory} & \textbf{Definition} & \textbf{Interviewee} \\
\hline
Insufficient Quality & AI-generated content is prone to errors and suffers from poor quality due to contextual limitations and the reuse of AI-generated information as training data. This leads to a lack of trust  & I01, I04, I06, I07, I09 \\
\hline
Trust considerations & Lack of trust or excessive trust in the use of AI & I01, I04, I06, I07, I09 \\
\hline
Data protection concerns & Risk of sensitive company and customer data being leaked or stolen during code analysis via external systems & I03, I05, I07, I08 \\
\hline
Higher energy and resource consumption than savings & The use of AI is perceived as disadvantageous if its own energy and resource consumption exceeds or relativizes the savings achievable through AI-supported optimizations. & I02, I03, I05, I06, I09, I10 \\
\hline
Fear of superiority & The fear that AI will become more intelligent than humans & I09 \\
\hline
\end{tabularx}
\end{table}

\subsection{Additional Remarks}
In addition to the requirements and barriers discussed above, one participant raised further remarks regarding the potential role of such tools in supporting sustainable software development. One recurring idea concerned the communication of general best practices. A participant noted that developers often learn primarily within the context of their own company or technology stack, which may limit their exposure to efficient programming techniques. In this regard, participants suggested that a tool could provide structured guidance, such as technology-specific recommendations or examples of efficient implementation patterns, to support developers in making more energy-efficient design decisions (I05).

Beyond concrete optimization guidance, one participant emphasized the importance of raising awareness for sustainability more broadly. A tool could help developers better understand how energy is consumed during software execution and what practical steps can be taken to reduce unnecessary resource usage (I10). This perspective was extended by suggestions that such tools could also highlight related aspects beyond code optimization, for example by informing developers about environmentally friendly infrastructure options such as green hosting or other opportunities for reducing environmental impact (I03).

Finally, a participant stressed that trust in the tool itself would be an important prerequisite for adoption. Concerns were expressed that automated optimizations might unintentionally introduce new errors or overlook edge cases, which could reduce confidence in the generated results. Ensuring transparency, reliability, and verifiability of optimization suggestions was therefore considered essential for building trust among developers (I09).

%% file: chapter/05_discussion.tex
\section{Discussion} \label{discussion}

This study investigated how software developers perceive energy efficiency in practice and which requirements and barriers shape the adoption of AI-assisted tools for sustainable software development. While prior research has mainly focused on technical optimization techniques and general barriers to energy-efficient software development, less attention has been paid to how energy-efficiency tools align with developers’ everyday workflows. Consequently, such tools are often difficult to introduce or remain unused in practice. The findings presented here address this gap by identifying developers’ perspectives on energy efficiency as well as the requirements and barriers that influence the adoption of energy efficiency optimization tools in daily development work.

Taken together, the findings highlight that, among other aspects, the effectiveness of energy efficiency approaches in software development is not determined solely by technical capabilities, but also by how well they fit existing development practices and organizational realities.

\subsection{Energy Efficiency in Software Development Practice}
A central finding of this study is that energy efficiency currently plays only a marginal and implicit role in everyday software development practice. Interviewees described energy efficiency not as an explicit development objective, but rather as a secondary outcome of established priorities such as functionality, performance, and code quality. Optimization efforts are therefore primarily driven by performance considerations or general engineering practices, with energy efficiency emerging incidentally rather than intentionally.
This implicit treatment of energy efficiency can be linked to the practical development context described by participants. Daily work was characterized by strong time pressure caused by release deadlines and the dominance of functional requirements, leaving limited room for additional optimization activities perceived as non-essential. Sustainability-related optimizations were frequently considered technically complex and time-consuming, which competes with feature development and delivery constraints. As a result, developers tend to focus on efficiency improvements that directly support existing development goals, such as performance optimization or clean code practices. This observation aligns with prior research showing that developers tend to prioritize performance, maintainability, and delivery constraints over explicit energy-related concerns \cite{ardito_understanding_2015, wysocki_methods_2025, pereira_ranking_2021, lange_energy_2025, balanza-martinez_tactics_2024}.

In addition to time constraints, participants highlighted uncertainty regarding the economic value of sustainability efforts. Developers questioned whether energy efficiency improvements would be recognized by customers or management, particularly when such optimizations do not directly contribute to visible functionality. This perceived lack of organizational incentives may further reduce the motivation to integrate sustainability considerations into everyday development decisions. Together with the reported knowledge gaps regarding software energy consumption, these findings indicate that barriers to sustainable software development extend beyond technical challenges and are embedded in organizational priorities and developers’ decision-making processes. As a result, sustainability considerations are often deprioritized in everyday development work when they do not clearly align with economic incentives or delivery goals. Similar dynamics have been described in the literature as a “vicious cycle,” in which practitioners hesitate to adopt academic solutions due to perceived lack of industrial relevance, while researchers struggle to transfer their approaches into real-world development environments \cite{balanza-martinez_tactics_2024, guldner_development_2024}.

At the same time, the interviews reveal a relatively coherent conceptual understanding of sustainable software among practitioners. Participants predominantly framed sustainability in terms of resource efficiency, emphasizing reduced consumption of computational resources through efficient implementation, selective component usage, and optimized runtime behavior. Sustainability was further linked to long-term software quality, including maintainability, clear architectural structure, and comprehensive documentation. This perspective reflects research showing that high architectural quality and modular design can significantly reduce energy consumption, with reported reductions of up to 30\% \cite{salmikuukka_guidelines_2024}.
However, the findings also suggest a simplified mental model of software energy efficiency among practitioners. Several participants implicitly associated sustainability improvements with reduced execution time, reflecting a common assumption that runtime performance and energy consumption are directly correlated. Existing research challenges this assumption, demonstrating that execution time and energy consumption do not necessarily scale proportionally \cite{balanza-martinez_tactics_2024, pereira_ranking_2021}. This indicates a gap between academic knowledge and developers’ practical understanding of energy efficiency. In practice, this gap may limit the adoption of energy-aware development practices, as developers lack clear guidance on which design or implementation decisions have the most significant impact on software energy consumption. 

Taken together, these findings suggest that sustainable software development is currently interpreted by practitioners less as a distinct engineering objective and more as an extension of established software quality practices.

\subsection{Requirements for Integrating Energy-Efficiency Support into Development Workflows}
The adoption of energy efficiency support tools can be interpreted through the lens of the TAM, which emphasizes perceived usefulness and perceived ease of use as key determinants of technology adoption \cite{Davis1986TAM}. Many of the requirements expressed by participants correspond directly to these two dimensions. 

The findings highlight that developers perceive the usefulness of energy efficiency tools primarily in terms of their ability to provide actionable and transparent support during development. Participants emphasized the need for concrete optimization suggestions grounded in code analysis rather than abstract recommendations. Such suggestions were expected to be explainable and directly connected to the developer’s current implementation context. In addition, interviewees stressed the importance of making the effects of optimizations visible through measurable feedback, such as metrics, comparisons between code versions, or real-time visualizations of energy consumption. This shows that energy efficiency improvements need to be made understandable, justifiable and observable  in order to support development decisions and communicate their value within organizations. Prior research similarly emphasizes that energy-efficient software development should be closely linked to continuous monitoring of energy consumption at the hardware level \cite{georgiou_software_2020}. By feeding real consumption data back into the development process, such monitoring mechanisms can support iterative optimization and align energy-efficiency considerations with established agile development practices \cite{georgiou_software_2020}.

For requirements that correspond to the TAM dimension of perceived ease of use, participants expressed the importance of the seamless integration of such tools into existing development environments and workflows. Participants stressed that sustainability support must fit naturally into established workflows, particularly within IDEs or CI/CD pipelines, to avoid additional friction or workflow disruption. This requirement reflects a broader tendency identified in software engineering research, where developer tools are most effective when feedback is embedded directly into development processes \cite{kruglov_developing_2023, georgiou_software_2020}. 

Overall, these findings suggest that the successful adoption of energy-efficiency tools depends not only on their analytical capabilities but also on their ability to integrate seamlessly into existing development workflows while providing transparent and actionable feedback.

\subsection{Opportunities of AI for Energy-Efficient Software Development} 

Interviewees primarily perceived artificial intelligence as a means to increase development efficiency by automating time-consuming tasks such as performance testing or code analysis. Participants further emphasized AI’s potential to support developers in analyzing complex codebases and identifying optimization patterns that may be difficult to detect manually. From this perspective, AI is viewed less as a standalone optimization mechanism and more as a tool that assists developers in navigating the increasing complexity of modern software systems.

Interestingly, the interviews highlight a difference in how the value of AI-assisted optimization is framed. While developers primarily emphasize productivity gains and workflow efficiency, research literature tends to evaluate AI-based approaches mainly in terms of their potential to reduce software energy consumption. Empirical work suggests that specialized AI-based solutions can reduce energy consumption during code generation under certain conditions \cite{ilager_green-code_2025}. At the same time, studies show that large language models often remain below expert-level optimization quality when generating code improvements \cite{cappendijk_generating_2024}. This discrepancy suggests that practitioners and research communities frame the value of AI differently: developers primarily associate AI with productivity improvements, whereas academic studies predominantly assess its impact on software energy consumption.

Taken together, these findings indicate that AI is not perceived as a replacement for developer expertise but rather as an augmentation mechanism that extends human development capabilities while still requiring human oversight and contextual judgment.

\subsection{Barriers to the Adoption of AI-Assisted Sustainability Tools} 

Despite the perceived opportunities of AI-based support, the interviews also revealed several barriers to its adoption. A central concern expressed by participants relates to the sustainability paradox of AI itself. Developers questioned whether the computational resources required to train and operate AI models might outweigh the energy savings achieved through AI-assisted optimization. This concern is supported by recent research highlighting the substantial energy consumption associated with large-scale AI systems \cite{vartziotis_carbon_2024}. These findings suggest that the potential benefits of AI-assisted optimization need to be evaluated in relation to the energy consumption of the AI systems themselves. In this context, making the computational resources and energy use of AI visible may be necessary to assess whether AI-based optimizations lead to net sustainability benefits in practice.  

Additional barriers relate to trust, reliability, and data governance. Participants expressed reservations regarding the quality and reliability of AI-generated results, particularly in light of known issues such as hallucinations or inconsistent outputs. These concerns reinforce the perception that AI-generated suggestions require verification before being integrated into production code. Prior research similarly indicates that generative AI often prioritizes the correctness or popularity of solutions rather than energy efficiency, meaning that sustainability considerations are frequently absent in generated code \cite{sikand_generative_2024, vartziotis_carbon_2024, tuttle_can_2024}. This suggests that AI-assisted energy optimization tools should provide transparent explanations of their recommendations and enable developers to verify the reliability of generated suggestions before integrating them into their code.

Furthermore, participants highlighted concerns regarding data protection and regulatory compliance, particularly when proprietary source code must be analyzed by external AI systems. These concerns are consistent with broader discussions surrounding GDPR compliance and secure deployment of AI systems in industrial environments \cite{ilager_green-code_2025, manimegalai_energy_2024, tiwari_evaluating_2023}. This underscores the necessity for AI-assisted energy optimization tools to proactively tackle this issue, for instance by offering privacy-preserving deployment options, such as locally executed analysis, to ensure that sensitive development data remains under organizational control.

Overall, the findings indicate that the adoption of AI-assisted sustainability tools depends not only on their technical capabilities but also on addressing concerns regarding reliability, data protection, and the overall sustainability of AI technologies themselves.

\subsection{Limitations}

This study is subject to several limitations. First, the qualitative design and relatively small sample size limit the transferability of the findings. While the interviews provide in-depth insights into developer perspectives, they do not allow for statistical conclusions about the broader software engineering population. Furthermore, Interview duration varied between participants. In one case, the interview was comparatively short (11 minutes) because the participant had reviewed the interview questions in advance and was therefore able to provide concise and focused responses. These differences in interview length may have affected the level of detail obtained across participants.
Second, the study focuses exclusively on software developers; perspectives from managers, product owners, or organizational decision-makers were not included, although these roles may strongly influence adoption decisions. Third, participants were recruited through the authors’ professional networks, which may introduce sampling bias. Fourth, gender was collected as a demographic characteristic but was not used for comparative analysis because the imbalanced sample would not permit meaningful or anonymity-preserving interpretation. As a result, the findings not fully reflect the diversity of perspectives within the software development community.
Furthermore, the study was conducted primarily within a German context, meaning that regulatory awareness, organizational culture, and attitudes toward sustainability and data protection may differ in other regions. The researchers’ own backgrounds in software engineering and human–computer interaction may also have influenced the design of the interview guide and interpretation of the data. Additionally, interaction effects between interviewer and participants, including shared professional background, implicit expectations, and social dynamics, may have shaped the depth or direction of certain responses.

%% file: chapter/06_conclusion.tex
\section{Summary and Future Research} \label{summary}

As the energy footprint of ICT continues to grow, improving software energy efficiency has become increasingly important. However, the findings of this study suggest that the main challenge lies less in the absence of technical solutions than in their integration into everyday development practice. Energy efficiency is rarely treated as an explicit development objective and instead tends to emerge indirectly through established priorities such as performance, and workflow efficiency. Consequently, sustainable software development is most likely to succeed when energy awareness is embedded within existing development processes rather than introduced as an additional responsibility. To support this integration, developers require tools that provide clear and actionable feedback on energy consumption, offer understandable optimization guidance, continuously measure energy-related metrics, and integrate smoothly into established workflows. Within this context, AI-assisted approaches are perceived as promising but only conditionally acceptable, as their adoption depends on transparency, reliability, and clearly demonstrable net sustainability benefits.

Overall, the findings underline the importance of designing energy-aware development tools that integrate naturally into existing workflows. The requirements identified in this study provide a structured basis for the design of such support tools.

Future research should extend these findings through larger and more diverse samples, including organizational stakeholders beyond developers. Furthermore, since our research focused on a qualitative perspective, a quantitative approach could validate and further enhance these findings. Empirical evaluations of prototype tools developed within the European GreenCode project could further examine how identified requirements influence real-world adoption and development behavior. In addition, future research should investigate methods for transparently assessing the net sustainability impact of AI-assisted development tools, considering lifecycle energy costs alongside optimization benefits. Longitudinal studies may also help understand how energy awareness evolves once continuous feedback mechanisms are integrated into development workflows.